\documentclass[aps,prl,twocolumn,showpacs,showkeys,square,numbers,amssymb,amsmath,nobibnotes,nofootinbib]{revtex4}
\usepackage{docs}%
\usepackage{bm}%
\usepackage[colorlinks=true,linkcolor=red]{hyperref}
\usepackage{amsmath}
\usepackage{amsfonts}
\usepackage{amssymb}

\usepackage{graphicx}
\usepackage[sort&compress]{natbib}
\def\hf{\hat{f}}
\begin{document}

%\preprint{}

\title{Large Deviations of the Maximum Eigenvalue in Wishart Random Matrices}


\author{Pierpaolo Vivo}
\affiliation{School of Information Systems, Computing \&
Mathematics\\Brunel University, Uxbridge, Middlesex, UB8
3PH\\United Kingdom} \email[]{pierpaolo.vivo@brunel.ac.uk}

\author{Satya N. Majumdar}
\affiliation{Laboratoire de Physique Th\'{e}orique et Mod\`{e}les
Statistiques (UMR 8626 du CNRS), Universit\'{e} Paris-Sud,
B\^{a}timent 100, 91405 Orsay Cedex\\ France}

\author{Oriol Bohigas}
\affiliation{Laboratoire de Physique Th\'{e}orique et Mod\`{e}les
Statistiques, Universit\'{e} Paris-Sud, B\^{a}timent 100, 91405
Orsay Cedex\\ France}

%\thanks{}


\date{\today}

\begin{abstract}
We compute analytically the probability of large fluctuations to
the left of the mean of the largest eigenvalue in the Wishart
(Laguerre) ensemble of positive definite random matrices. We show
that the probability that all the eigenvalues of a ($N\times N $)
Wishart matrix $W=X^T X$ (where $X$ is a rectangular $M\times N$
matrix with independent Gaussian entries) are smaller than the
mean value $\langle\lambda\rangle=N/c$ decreases for large $N$ as
$\sim \exp\left[-\frac{\beta}{2}N^2
\Phi_{-}\left(\frac{2}{\sqrt{c}}+1;c\right)\right]$, where
$\beta=1,2$ corresponds respectively to real and complex Wishart
matrices, $c=N/M\le 1$ and $\Phi_{-}(x;c)$ is a rate (sometimes
also called large deviation) function that we compute explicitly.
The result for the Anti-Wishart case ($M<N$) simply follows by
exchanging $M$ and $N$. We also analytically determine the average
spectral density of an ensemble of Wishart matrices whose
eigenvalues are constrained to be smaller than a fixed barrier.
Numerical simulations are in excellent agreement with the
analytical predictions.
\end{abstract}


\maketitle


\section{Introduction}

Consider a rectangular $(M\times N)$ matrix $X$ whose elements
$X_{ij}$ represent some data. The $N$ entries of each of the $M$
rows constitute the components of an $N$-dimensional vector $\vec
X_i$ (with $i=1,2,\ldots, M$). The vector $\vec X_i$ (the $i$-th
row of the array) represents the $i$-th sample of the data and the
matrix element $X_{ij}$ represents the $j$-th component of the
vector ${\vec X_i}$. For example, suppose we are considering a
population of $M$ students in a class, and for each student we
have the data of their heights, their marks in an examination,
their weights etc. forming a vector of $N$ elements (or traits)
for each of the $M$ students. Then the product $W=X^TX$ is a
positive definite symmetric $(N\times N)$ matrix that represents the
covariance
matrix of the data (unnormalized). This matrix characterizes the
correlations between different traits. The spectral properties of
this matrix, i.e., its eigenvectors and eigenvalues, play a very
important role in the so called `principal components analysis'
(PCA) of multivariate data, a technique that is used regularly in
detecting hidden patterns in data and also in image
processing~\cite{Wilks,Fukunaga,Smith}, amongst other
applications. In PCA, one diagonalizes the covariance matrix $W$
and identifies all the eigenvalues and eigenvectors. The data are
usually maximally scattered in the direction of its principal
eigenvector, corresponding to the largest eigenvalue and are least
scattered in the direction of the eigenvector corresponding to the
minimum eigenvalue. One can then prune the data by successively
getting rid of the components (setting them to zero) along the
eigenvectors corresponding to the smaller eigenvalues, but
retaining the components along the larger eigenvalues, in
particular those corresponding to the maximal eigenvalue. This
method thus reduces the effective dimension of the data. This
technique is called `dimensional reduction' and forms the basis of
e.g, image compression in computer vision~\cite{Smith}.

When the underlying data are random, e.g. the elements of the
matrix $X$ are independent and identically distributed (i.i.d)
random variables, real or complex, drawn from a Gaussian
distribution, the product matrices $W=X^\dagger X$ constitute the
so called Wishart ensemble, named after Wishart who first
introduced them~\cite{Wishart}. In literature one can also find
the term 'Laguerre' ensemble, because the Laguerre polynomials
arise in the analytical treatment of its spectral properties.

These Wishart random matrices have been extremely useful in
multivariate statistical data analysis~\cite{Wilks,Johnstone}
mentioned above (where $W$ represents the covariance matrix) with
applications in various fields ranging from meteorological
data~\cite{Preisendorfer} to finance~\cite{BP,Burda}. Such
matrices are also useful to analyze the capacity of channels with
multiple antennae and receivers~\cite{SP}. They also appear in
nuclear physics~\cite{Fyo1}, quantum chromodynamics~\cite{QCD} and
also in statistical physics such as in a class of
$(1+1)$-dimensional directed polymer problems~\cite{Johansson}.
Recently, Wishart matrices have also been used in the context of
knowledge networks~\cite{MZ1} and new mathematical results for the
distribution of the matrix elements for the Anti-Wishart matrices
(when $M<N$) have been obtained~\cite{Z2,JN}.

\begin{figure}
\includegraphics[width=.9\hsize]{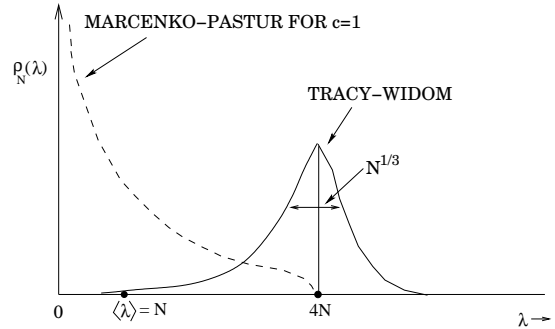}
\caption{The dashed line shows schematically the Mar\v{c}enko-Pastur
form of the average density of states for $c=1$. The average
eigenvalue for $c=1$ is $\langle \lambda \rangle=N$. For $c=1$,
the largest eigenvalue is centered around its mean $\langle
\lambda_{\rm max} \rangle= 4N$ and fluctuates over a scale of
width $N^{1/3}$. The probability of fluctuations on this scale is
described by the Tracy-Widom distribution (shown schematically).}
\label{fig:mp}
\end{figure}

Given that the joint distribution of the elements of the $(M\times
N)$ matrix $X$ (real or complex) is a Gaussian, $P[X]\propto
\exp\left[-\frac{\beta}{2}\, {\rm tr}(X^\dagger X)\right]$ where
the Dyson index $\beta=1,2$ corresponds respectively to real and
complex matrices \cite{Dys:new}, the spectral properties of the
Wishart matrix $W=X^\dagger X$ have been studied extensively for
many decades. For the case when $M\ge N$ (the number of samples is
larger than the dimension) it is known that all the eigenvalues
are positive, a typical eigenvalue scales as $\lambda\sim N$ for
large $N$, and the average density of eigenvalues in the large $N$
limit has a scaling form $\rho_N(\lambda)\approx
\frac{1}{N}f\left(\frac{\lambda}{N}\right)$, where $f(x)$ is the
Mar\v{c}enko-Pastur~\cite{MP} function on the compact support $x\in
[x_{-},x_{+}]$:
\begin{equation}
f(x)= \frac{1}{2\pi x}\, \sqrt{(x_{+}-x)(x-x_{-})}
\label{Marcenko-Pastur}
\end{equation}
with $x_{\pm}=\left(\frac{1}{\sqrt{c}}\pm 1\right)^2$ and $c=N/M$
(with $c\le 1$). (This result was also rederived by a
different method by Dyson~\cite{Dyson} and the spectral fluctuations were
numerically investigated by Bohigas et al. \cite{Bohigas}). Thus, for
$c\le
1$, all the eigenvalues lie within a compact Mar\v{c}enko-Pastur sea and
the
average eigenvalue,
\begin{equation}
\langle \lambda \rangle = \int_0^{\infty} \rho_N(\lambda) \lambda
d\lambda=\frac{N}{c}. \label{avl}
\end{equation}
For all $c<1$, the distribution goes to zero at the edges $x_{-}$
and $x_{+}$. For the case $c=1$ ($x_{-}=0$ and $x_{+}=4$), the
distribution diverges as
$x^{-1/2}$ at the origin, $f(x)= \frac{1}{2\pi}\sqrt{(4-x)/x}$
for $0\le x\le 4$ (shown schematically in Fig. \ref{fig:mp}). For
the Anti-Wishart case ($M<N$, i.e., $c>1$) where one has $M$
positive eigenvalues (the rest of the $(N-M)$ eigenvalues are
identically zero), the corresponding result can be obtained from
the $M\ge N$ case simply by exchanging $M$ and $N$.

Another important issue in the context of PCA is the distribution
of the largest eigenvalue of a Wishart random matrix and a lot of
recent work has been devoted to this
question~\cite{Edelman,Forrester,Johansson,Johnstone, DE,ES}. From
the exact analytical form of the density of states, it follows
that the average of the maximum eigenvalue for large $N$ is
$\langle \lambda_{\rm max}\rangle \approx x_{+}(c)\,N$ where
$x_{+}(c)=\left(\frac{1}{\sqrt c}+1\right)^2$. However, for finite
but large $N$, the maximum eigenvalue fluctuates, around its mean
$x_{+}(c)\,N$, from one sample to another. A natural question is:
what is the full probability distribution of the largest
eigenvalue $\lambda_{\rm max}$? Recently,
Johansson~\cite{Johansson} and independently
Johnstone~\cite{Johnstone} showed that for large $N$ these
fluctuations {\em typically} occur over a scale $\sim O(N^{1/3})$
around the mean, i.e. the upper edge of the Mar\v{c}enko-Pastur
distribution, and the probability of {\em typical} fluctuations
$\chi=N^{-1/3}[\lambda_{\rm max}-x_{+}(c)N]$, properly centered
and scaled, is described by the well known Tracy-Widom
distribution (see Section \ref{Section II} for details).

Note that the Tracy-Widom distribution describes the probability
of {\em typical and small} fluctuations of $\lambda_{\rm max}$
over a narrow region of width $\sim O(N^{1/3})$ around the mean
$\langle \lambda_{\rm max}\rangle \approx x_{+}(c) \,N $. A
question that is particularly important in the context of PCA is
how to describe the probability of {\em atypical and large}
fluctuations of $\lambda_{\rm max}$ around its mean, say over a
wider region of width $\sim O(N)$? For example, what is the
probability that all the eigenvalues of a Wishart random matrix
are less than the average $\langle \lambda\rangle \approx N/c$ for
large N? This is the same as the probability that $\lambda_{\rm
max}\le N/c$. Since $\langle \lambda_{\rm max}\rangle \approx
x_{+}(c)\,N $, this requires the computation of the probability of
an extremely rare event characterizing a large deviation of $\sim
O(N)$ to the left of the mean (see e.g. a schematic picture for
$c=1$ in Fig. \ref{fig:mp}). Questions of this kind have been
recently addressed in ref. \cite{DM} on which we heavily rely,
while for the general large deviations theory in connection with
random matrices the reader is referred to \cite{Hiai}.

In the context of PCA, this \emph{large deviation} issue arises
quite naturally because one is there interested in getting rid of
redundant data by the `dimension reduction' technique and keeping
only the principal part of the data in the direction of the
eigenvector representing the maximum eigenvalue, as mentioned
before. The `dimension reduction' technique works efficiently only
if the largest eigenvalue is much larger than the other
eigenvalues. However, if the largest eigenvalue is comparable to
the average eigenvalue $\langle \lambda \rangle$, the PCA
technique is not very useful. Thus, the knowledge of large
negative fluctuations of $\lambda_{\rm max}$ from its mean
$\langle \lambda_{\rm max}\rangle \approx x_{+}(c)\,N $ provides
useful information about the efficiency of the PCA technique.

The main purpose of this paper is to provide exact analytical
results for these large negative fluctuations of $\lambda_{\rm
max}$ from its mean value. Rigorous mathematical results about the
asymptotics of the Airy-kernel determinant (i.e. the probability
that the largest eigenvalue lies deep inside the
Mar\v{c}enko-Pastur sea) for the case $c=1$ and $\beta=2$ have
been recently obtained \cite{Deift}. Here we follow the Coulomb
gas approach \cite{Dys:new}\cite{Mehta}, which interprets the
eigenvalues of a random matrix as a fluid of charged interacting
particles, and use standard functional integration methods of
statistical physics. This approach has been exploited in the
context of the Laguerre ensemble for the first time by Chen and
Manning \cite{Chen}, who performed a detailed asymptotic analysis
of the level spacing for general $\beta$ and $\alpha>-1$ (where
$\alpha$ is essentially the prefactor of the external logarithmic
potential, see e.g. \eqref{Stationarity2}) and determined the
distribution of the two smallest eigenvalues in a certain
double-scaling limit. These techniques have been also recently
used to obtain analytically the large negative fluctuations of the
maximum eigenvalue for the Gaussian ensembles~\cite{DM}. Here we
adopt this method for the Wishart ensemble.

We show that for $c\le 1$, the probability of large fluctuations
to the left of the mean $\langle \lambda_{\rm max}\rangle \approx
x_{+}(c)\,N$ behaves, for large $N$, as
\begin{align}
\nonumber {\rm Prob}\left[\lambda_{\rm max}\le t, N\right] &\sim \\
&\exp\left[{-\frac{\beta}{2} N^2
\Phi_{-}\left(\frac{x_{+}(c)\,N-t}{N};\,c\right)}\right]
\label{ldf1}
\end{align}
where $t\sim O(N)\le x_{+}(c)\, N$ is located deep inside the
Mar\v{c}enko-Pastur sea and $\Phi_{-}(x;c)$ is a certain {\em
left} rate (sometimes also called large deviation) function with
$x$ being the main argument of the function and $c$ being a
parameter. In this paper, we compute the rate function
$\Phi_{-}(x;c)$ explicitly. Knowing this function, it then follows
that for large $N$
\begin{equation}
{\rm Prob}\left[\lambda_{\rm max}\le \langle \lambda\rangle=N/c,
N\right]\sim \exp(-\theta(c) N^2), \label{exp1}
\end{equation}
where the coefficient
\begin{equation}
\theta(c)= \frac{\beta}{2} \Phi_{-}\left(\frac{2}{\sqrt
c}+1;\,c\right). \label{theta1}
\end{equation}
For example, for the case $c=1$ ($M=N$), we show that
\begin{equation}
\theta(1) = {\beta}\,\left(\log 2-\frac{33}{64}\right)=
0.177522\dots \, \beta. \label{theta2}
\end{equation}
The corresponding result for the Anti-Wishart matrices $(M<N)$
simply follows by exchanging $M$ and $N$. In this paper, we focus
only on the {\em left} large deviations of $\lambda_{\rm max}$.
The corresponding probability of large fluctuations of
$\lambda_{\rm max}$ to the {\em right} of the mean $\langle
\lambda_{\rm max} \rangle$ was previously computed explicitly by
Johansson~\cite{Johansson} (see the next section for details).

As a byproduct of our analysis, we provide the general expression
for the spectral density of a constrained Wishart ensemble of
matrices whose eigenvalues are restricted to be smaller than a
fixed barrier.

The paper is organized as follows. In section \ref{Section II}, we
set up notations, we provide some mathematical preliminaries and
we recall some known results for the Wishart random matrices as
well as, for the sake of comparison, of Gaussian random matrices.
In section \ref{Section III} we outline the functional integration
method followed by the steepest descent calculation. In subsection
\ref{Subsection IIIa}, we derive the left rate function explicitly
for the special case $c=1$ and in subsection \ref{Subsection IIIb}
we extend the results to the case $c< 1$. In section \ref{Section
IV} numerical simulations are compared to the analytical
predictions. Section \ref{Section V} concludes the paper with a
summary and discussion, while the analytical computation of the
rate function for $c<1$ is reported in the Appendix.

\section{ Wishart and Gaussian Random Matrices: Some
Preliminaries}\label{Section II}

We consider a rectangular $(M\times N)$ matrix $X$ with $M$ rows
(representing $M$ different samples) and $N$ columns (representing
$N$ components of each sample). We assume that the entries of the
matrix $X$ are i.i.d random variables each drawn independently
from a standard normal distribution, such that the joint
distribution of the elements is given by $P[X]\propto
\exp\left[-\frac{\beta}{2}\, {\rm tr}(X^\dagger X)\right]$ where
the Dyson index $\beta=1,2$ corresponds respectively to real and
complex matrices \cite{Dys:new}. One then constructs the Wishart
matrix $W=X^{\dagger} X$ by taking the product. The first natural
question is: Given the distribution of $X$, what is the joint
distribution of the elements of $W$? It turns out that this is not
quite easy to compute. For the case when $M\ge N$ (when the number
of samples is larger or equal to the dimension of the vector),
this was computed by Wishart~\cite{Wishart}. The corresponding
calculation for the opposite `Anti-Wishart' case, when $M<N$,
turns out to be much more complicated. This was first obtained
numerically~\cite{MZ1} and only recently an analytical expression
has been found~\cite{Z2, JN}.

In contrast with the probability distribution of the matrix
elements of $W$ itself, the joint probability distribution (jpd)
of its eigenvalues was known since a long time~\cite{James}, and
from it all the interesting spectral properties of the ensemble
can be derived. We summarize them here together with the
corresponding ones for the Gaussian ensemble.

\subsection{Wishart (Anti-Wishart) ensemble}

For the case when $M\ge N$, all the $N$ eigenvalues are positive
and their jpd is given by
\begin{align}\label{jpdWishart}
  \nonumber P_N(\lambda_1,\ldots,\lambda_N) &=K_N e^{-\frac{\beta}{2}\sum_{i=1}^N\lambda_i}
  \prod_{i=1}^N
  \lambda_i^{\frac{\beta}{2}(1+M-N)-1}\times\\
  &\times\prod_{j<k}|\lambda_j-\lambda_k|^\beta
\end{align}
where $K_N$ is a normalization constant and the parameter
$\beta=1,2$ corresponds respectively to the real and complex $X$.
On the other hand, for the Anti-Wishart case ($M<N$), there are
only $M$ positive eigenvalues (the rest of the $N-M$ eigenvalues
are exactly $0$) and their jpd is given exactly by the same
formula as in \eqref{jpdWishart} except that $N$ and $M$ are
interchanged~\cite{JN}.

For the Wishart matrices with $M\ge N$, in the large $N$ limit,
the average density of states has the scaling form,
$\rho_N(\lambda)\approx
\frac{1}{N}f\left(\frac{\lambda}{N}\right)$ where $f(x)$ is the
Mar\v{c}enko-Pastur~\cite{MP} function defined in
\eqref{Marcenko-Pastur}. The corresponding result for the
Anti-Wishart case ($M<N$) where one has $M$ eigenvalues, simply
follows by exchanging $M$ and $N$.

For large $N$ the maximum eigenvalue fluctuates around its average
$\langle \lambda_{\rm max}\rangle \approx x_{+}(c) N$ and the
typical fluctuation occurs over a scale of width $O(N^{1/3})$
around the mean. Johansson~\cite{Johansson} and independently
Johnstone~\cite{Johnstone} computed the limiting distribution of
these {\em typical} fluctuations around the mean. They showed that
for large $N$ and for $c\le 1$~\cite{Johansson,Johnstone}
\begin{equation}
\lambda_{\rm max} = \left(\frac{1}{\sqrt{c}}+ 1\right)^2\,N +
c^{1/6} \left(\frac{1}{\sqrt{c}}+ 1\right)^{4/3}\, N^{1/3} \chi
\label{JJ}
\end{equation}
where the random variable $\chi$ has an $N$-independent limiting
distribution ${\rm Prob}(\chi\le x)= F_{\beta}(x)$, which is the
well known Tracy-Widom distribution (see below).

\subsection{Gaussian ensemble}
In the case of a random $(N\times N)$ Gaussian
matrix~\cite{Wigner,Mehta}, the jpd of eigenvalues is given by:
\begin{align}
P_N(\lambda_1,\ldots,\lambda_N) &= B_N\,
e^{-\frac{\beta}{2}\sum_{i=1}^N\lambda_i^2}\,\prod_{j<k}|\lambda_j-\lambda_k|^\beta
\label{pdfgaussian}
\end{align}
where $B_N$ normalizes the jpd and $\beta=1$, $2$ and $4$
correspond respectively to the GOE (Gaussian orthogonal ensemble),
GUE (Gaussian unitary ensemble) and GSE (Gaussian symplectic
ensemble).

The average density of states in the large $N$ limit has the
scaling form $\rho_N(\lambda)\approx \frac{1}{\sqrt{N}}f_{\rm
sc}\left(\frac{\lambda}{\sqrt{N}}\right)$ where $f_{\rm sc}(x)$ is
the famous Wigner semi-circular law: $f_{\rm sc}(x)=
\sqrt{\frac{1}{\pi}\,[2-x^2]}$ with compact support over $x\in
[-\sqrt{2},\sqrt{2}]$.

Furthermore, the analogous asymptotic form of $\lambda_{\rm max}$
is known to be~\cite{TW1}
\begin{equation}
\lambda_{\rm max} = \sqrt{2N} + \frac{N^{-1/6}}{\sqrt{2}}\, \chi
\label{tw2}
\end{equation}
where the random variable $\chi$ has again the limiting
$N$-independent distribution, ${\rm Prob}[\chi \le x] =
F_{\beta}(x)$. \vspace{5pt}

In this paper, the main interest is focused on the largest
eigenvalue. In summary, the \emph{scaled variables} $\lambda_{\rm
max}/N$ in the Wishart case and $\lambda_{\rm max}/\sqrt{N}$ in
the Gaussian case, both {\em typically} fluctuate over a region of
width $\sim O(N^{-2/3})$ around their mean and these typical
fluctuations are described by the Tracy-Widom law $F_\beta(x)$.

The function $F_{\beta}(x)$, computed as a solution of a nonlinear
Painlev\'{e} differential equation~\cite{TW1}, approaches to $1$
as $x\to \infty$ and decays rapidly to zero as $x\to -\infty$. For
example, for $\beta=2$, $F_2(x)$ has the following
tails~\cite{TW1},
\begin{eqnarray}
F_2(x) &\to & 1- O\left(\exp[-4x^{3/2}/3]\right)\quad\, {\rm
as}\,\,\, x\to \infty
\nonumber \\
&\to & \exp[-|x|^3/12] \quad\, {\rm as}\,\,\, x\to -\infty.
\label{asymp1}
\end{eqnarray}
The probability density function $f_{\beta}(x)=dF_{\beta}/dx$ thus
has highly asymmetric tails.

It follows from \eqref{JJ} that in the Wishart case, the
Tracy-Widom distribution describes the probability of {\em typical
and small} fluctuations of $\lambda_{\rm max}$ over a narrow
region of width $\sim O(N^{1/3})$ around the mean $\langle
\lambda_{\rm max}\rangle \approx x_{+}(c) \,N $ where
$x_{+}(c)=\left(\frac{1}{\sqrt c}+1\right)^2$.

As mentioned in the introduction, in this paper we are concerned
not with the {\em typical small} fluctuations of $O(N^{1/3})$
around the mean, but rather with {\em atypical large} fluctuations
of $O(N)$. Thus, we are interested in computing the probability of
extremely rare events. In fact, the question about the large
deviation of the largest eigenvalue was addressed before in
\cite{Johansson} and it was proved by rigorous methods that for
$c\le 1$ the probability of {\em large} fluctuations to the left
of the mean $\langle \lambda_{\rm max}\rangle \approx x_{+}(c)\,N
$, behaves for large $N$ as in \eqref{ldf1}, but an explicit
expression for the left rate function $\Phi_{-}(x;c)$ was missing
so far. On the other hand, for {\em large} fluctuations to the
right of the mean $\langle \lambda_{\rm max}\rangle \approx
x_{+}(c)\,N $,
\begin{align}
\nonumber
1-{\rm Prob}\left[\lambda_{\rm max}\le t, N\right] &\sim  \\
& \exp\left[{-\frac{\beta}{2}N
\Phi_{+}\left(\frac{t-x_{+}(c)\,N}{N};\,c\right)}\right]
\label{ldf2}
\end{align}
for $t\sim O(N)\ge x_{+}(c)\,N$ located outside the
Mar\v{c}enko-Pastur sea to its right and $\Phi_{+}(x;c)$ is the
{\em right} rate function that was obtained explicitly in
\cite{Johansson}.

%\\
%\\
% \emph{ From the joint pdf, one can calculate the
%average density of states defined as $\rho_N(\lambda)=
%\sum_{i=1}^N\langle \delta(\lambda-\lambda_i)\rangle/N$, which
%counts the average number of eigenvalues between $\lambda$ and
%$\lambda + d\lambda$ per unit length. $\rho_N(\lambda)$ is simply
%the marginal of the joint pdf in \eqref{jpdWishart}:
%\begin{eqnarray}
%\rho_N(\lambda) &=& \langle \frac{1}{N} \sum_{i=1}^N \delta(\lambda-\lambda_i)\rangle \nonumber \\
%&=& \int_{\infty}^{\infty} d\lambda_2\dots d\lambda_N P_N(\lambda,
%\lambda_2,\dots,\lambda_N). \label{marginal}
%\end{eqnarray}

%\\
%\\
%\emph{In our case, we are interested in computing explicitly the
%left large deviation function, since knowing this function we can
%then compute the probability that all eigenvalues are less than
%the average $\langle \lambda
%\rangle=N/c$ as mentioned in \eqref{exp1} of the introduction.}\\
%\\
The purpose of this paper is to provide an exact result for
$\Phi_{-}(x;c)$ for all $c\le 1$. For $c>1$ (Anti-Wishart) case,
the result holds with $M$ and $N$ interchanged. Let us summarize
our main results. For the case $c=1$, we give an explicit
expression for the left rate function $\Phi_{-}(x;1)$ as stated in
\eqref{Large deviation function}. Subsequently, the results in
\eqref{theta1} and \eqref{theta2} follow. For $c< 1$, the function
$\Phi_{-}(x;c)$ has a rather long analytical expression which is
derived in the Appendix. However, the function can be easily
evaluated using Mathematica$^\circledR$ as illustrated in Fig.
\ref{Phicmin1}.

These results should be compared to the corresponding ones for the
Gaussian case. For the Gaussian ensemble, the left large
deviations follow a similar law, namely
\begin{align}
\nonumber
{\rm Prob}\left[\lambda_{\rm max}\le t, N\right] &\sim \\
&\exp\left[{-\beta N^2 \Phi_{-}^{Gauss}\left(
\frac{\sqrt{2N}-t}{\sqrt{N}}\right)}\right] \label{ldfg1}
\end{align}
where $t\sim O(N^{1/2})\le \sqrt{2N}$ is located deep inside the
Wigner sea. For the Gaussian case, $\langle \lambda\rangle=0$.
Thus, the corresponding question about the probability that all
eigenvalues are less than their average is the same as the
probability that all eigenvalues are negative. This probability
plays a very important role in determining the average number of
maxima of a random smooth potential, where a stationary point is a
local maximum if all the eigenvalues of the associated hessian
matrix are negative. The calculation of this probability has been
a subject of many theoretical and numerical studies with important
applications in disordered systems, supercooled liquids, glassy
models~\cite{CGG,Fyodorov} and more recently in anthropic
principle based string theory~\cite{Susskind,MH,AE}. Very
recently, the left rate function $\Phi_{-}^{Gauss}(y)$ has been
computed exactly using functional integration methods~\cite{DM}.
Using this result, it was shown in \cite{DM} that for Gaussian
matrices,
\begin{equation}
{\rm Prob}\left[\lambda_{\rm max}\le 0; N\right]\sim \exp(-\beta
\theta N^2), \label{exp11}
\end{equation}
for large $N$ where the coefficient
\begin{equation}
\theta= \frac{1}{4}\log 3=0.274653\dots \label{thetag}
\end{equation}
In this paper, we adapt the techniques used in \cite{DM} for
Gaussian matrices to the Wishart case. Similar techniques have
recently been used also in other problems such as in the
calculation of the average number of stationary points for a
Gaussian random field with $N$ components in the large $N$
limit~\cite{BrayDean,FSW}.

Incidentally, let us remark that our problem might be tackled also
from the completely different viewpoint of zero-dimensional
replica field theories thanks to their recently discovered exact
integrability \cite{Kanz}. This yet unexplored route may provide
an independent derivation of our results.

\section{Functional Integration and the method of Steepest
Descent}\label{Section III}

Our starting point is the joint distribution of eigenvalues of the
Wishart ensemble in \eqref{jpdWishart}. Let $P_N(t)$ be the
probability that the maximum eigenvalue $\lambda_{\rm max}$ is
less than or equal to $t$. Clearly, this is also the probability
that all the eigenvalues are less than or equal to $t$ and can be
expressed as a ratio of two multiple integrals
\begin{align}\label{P_N(t)}
  \nonumber P_N(t) &=\mathrm{Prob}[\lambda_{\mathrm{max}}\leq
  t]=\frac{Z_1(t)}{Z_0}=\\
&=\frac{\int_0^{t}\ldots\int_0^{t} d\lambda_1\ldots
d\lambda_N\exp(-\frac{\beta}{2}F[\vec{\lambda}])}{\int_0^\infty\ldots\int_0^\infty
d\lambda_1\ldots d\lambda_N\exp(-\frac{\beta}{2}F[\vec{\lambda}])}
\end{align}
where:
\begin{equation}\label{F}
F[\vec{\lambda}]=\sum_{i=1}^N\lambda_i-(1+M-N-\frac{2}{\beta})\sum_{i=1}^N
\log\lambda_i-\sum_{j\neq k}\log|\lambda_j-\lambda_k|
\end{equation}
Written in this form, $F$ mimics the free energy of a $2$-d
Coulomb gas of interacting particles confined to the positive
half-line ($\lambda>0$) and subject to an external
linear+logarithmic potential, as mentioned in the introduction.
The denominator in \eqref{P_N(t)}, which is simply a normalization
constant, represents the partition function of a free or
`unconstrained' Coulomb gas over $\lambda\in [0,\infty)$. The
numerator, on the other hand, represents the partition function of
the same Coulomb gas, but with the additional constraint that the
gas is confined inside the box $\lambda\in [0,t]$, i.e., there is
an additional wall or infinite barrier at the position
$\lambda=t$. We will refer to the numerator as the partition
function of a `constrained' Coulomb gas. The constraint should not
be effective when $t<x_{-}$ or $t>x_{+}$.

Note that in the Gaussian case, the external potential is harmonic
over the whole real line ($V(\lambda)= \lambda^2/2$), while in the
Wishart case, $V(\lambda)=\infty$ for $\lambda<0$ (infinite
barrier at $\lambda=0$) and $V(\lambda)=\lambda -
(1+M-N-2/\beta)\log \lambda$ for $\lambda>0$ representing a
linear+logarithmic potential. By comparing the external potential
and the logarithmic interaction term, it is easy to see that while
for Gaussian ensembles a typical eigenvalue scales as $\lambda\sim
\sqrt{N}$ for large $N$, for the Wishart case it scales as
$\lambda\sim N$.

After defining the \emph{constrained charge density}:
\begin{equation}\label{SpatialDensityOfCharge}
  \hat{\varrho}_N(\lambda):=\varrho_N(\lambda;t)=\frac{1}{N}\sum_{i=1}^N \delta(\lambda-\lambda_i)
  \theta(t-\lambda)
\end{equation}
and taking into account the following trivial identity for a
generic function $h(x)$:
\begin{equation}\label{Identity}
\sum_{i=1}^N h(\lambda_i)=N\int
d\lambda\hat\varrho_N(\lambda)h(\lambda)
\end{equation}
we may express, for large $N$, the partition function $Z_1(t)$ in
\eqref{P_N(t)} as a functional integral~\cite{DM}:
\begin{align}\label{action}
\nonumber Z_1(t) \propto
&\int\mathcal{D}[\hat\varrho_N]\exp\left\{-\frac{\beta}{2}\right.
\left[N\int_0^{t}
\hat\varrho_N(\lambda)\lambda d\lambda \right.\\
\nonumber &-N(M-N+1-2/\beta) \int_0^{t}
\hat\varrho_N(\lambda)\log\lambda d\lambda\\
\nonumber
&-N^2\int_0^{t}\int_0^{t}\hat\varrho_N(\lambda)\hat\varrho_N(\lambda^\prime)
\log|\lambda-\lambda^\prime|d\lambda d\lambda^\prime\\
&-N\int_0^{t} \hat\varrho_N(\lambda)\log[\hat\varrho_N(\lambda)]
d\lambda\left.\right]\left.\right\}
\end{align}
where the last entropic term is of order $O(N)$ and arises from
the change of variables in going from an ordinary multiple
integral to a functional integral, $[\{\lambda_i\}]\to
[\hat{\varrho}_N(\lambda)]$. The constrained charge density
$\hat\varrho_N(\lambda)$ satisfies the obvious constraints
$\hat\varrho_N(\lambda)=0$ for $\lambda>t$ and
$\int_0^{t}\hat\varrho_N(\lambda) d\lambda =1$.

Since we are interested in fluctuations of $\sim O(N)$, it is
convenient to work with the rescaled variables $\lambda = x N $
and  $\zeta=t/N$. It is also reasonable to assume that for large $N$, the
charge
density scales accordingly as
$\hat\varrho_N(\lambda)=N^{-1}\hf(\lambda/N)$, so that
$\hf(x)=0$ for $x>\zeta$ and $\int_0^\zeta \hf(x)dx=1$.

In terms of the rescaled variables, the energy term in
\eqref{action} becomes proportional to $N^2$ while the entropy
term ($\sim O(N)$) is subdominant in the large $N$ limit.
Eventually we can write:
\begin{equation}\label{Eventual action}
  Z_1(\zeta)\propto\int\mathcal{D}[\hf]\exp\left(-\frac{\beta}{2}N^2 S[\hf(x);\zeta]+O(N)\right)
\end{equation}
where:
\begin{align}\label{S}
        \nonumber  S[\hf(x);\zeta] &=\int_0^\zeta x \hf(x)dx-\alpha\int_0^\zeta \hf(x)\log(x)dx+\\
    \nonumber &-\int_0^\zeta\int_0^\zeta \hf(x)\hf(x^\prime)\log|x-x^\prime|dx
    dx^\prime+\\
    &+C_1\left[\int_0^\zeta \hf(x)dx-1\right]
\end{align}
where we have introduced the parameter $\alpha=\frac{1-c}{c}$ for
later convenience. In \eqref{S}, $C_1$ is a Lagrange multiplier
enforcing the normalization of $\hf$.

For large $N$ we can evaluate the leading contribution to the
action \eqref{S} by the method of steepest descent. This gives:
\begin{equation}\label{Steep}
  Z_1(\zeta)\propto\exp\left[-\frac{\beta}{2}N^2 S[\hf^\star(x);\zeta]+O(N)\right]
\end{equation}
where $\hf^\star$ is the solution of the stationarity condition:
\begin{align}\label{Stationarity}
  \frac{\delta S[\hf(x);\zeta] }{\delta \hf(x)}=0
\end{align}
This gives for $0\leq x\leq \zeta$:
\begin{equation}\label{Stationarity2}
  x-\alpha\log x+C_1=2\int_0^\zeta \hf(x^\prime)\log|x-x^\prime|dx^\prime
\end{equation}
Differentiating \eqref{Stationarity2} once with respect to $x$
gives:
\begin{equation}\label{Stationarity3}
  \frac{1}{2}-\frac{\alpha}{2x}=\mathcal{P}\int_0^\zeta\frac{\hf(x^\prime)}{x-x^\prime}dx^\prime\qquad 0\leq x\leq
  \zeta
\end{equation}
where $\mathcal{P}$ denotes the Cauchy principal part.

Finding a solution to the integral equation \eqref{Stationarity3}
is the main technical task. The next two subsections are devoted
to the solution of \eqref{Stationarity3}, first for the special
case $c=1$ and then for $0<c<1$. We remark that the solution of
\eqref{Stationarity3} gives the average density of eigenvalues in
the limit of large $N$ for an ensemble of Wishart matrices whose
rescaled eigenvalues are restricted to be smaller than the barrier
$\zeta$. We will refer to $\hf(x)$ as the \emph{constrained}
spectral density.

Before proceeding to the technical points, it may be informative
to first summarize the results for the constrained spectral
density $\hf(x)$ in the general $0<c\leq 1$ case. The most general
form is:
\begin{equation}\label{Generic form}
    \hf(x)=\frac{1}{2\pi}
\frac{\sqrt{x-L_1(c,\zeta)}}
{\sqrt{\zeta-x}}\left[\frac{A(c,\zeta)-x}{x}\right]
\end{equation}
where $L_1$ is the lower edge of the spectrum and $A$ is related
to the mutual position of the barrier with respect to the lower
edge. In the following table, we schematically anticipate the
values for $L_1$ and $A$ in the different regions of the
$(c,\zeta)$ plane:
\begin{table}[htb]
\begin{tabular}{|c|c|c|}
\hline & $c=1$ & $0<c<1$\\
\hline $0<\zeta<x_{+}$ & $L_1=0$ \eqref{Constrained density} & $L_1$: see \eqref{L1}\\
\cline{2-3} (barrier \emph{effective}) & $A=\frac{\zeta+4}{2}$ \eqref{Constrained density} & $A=\alpha\sqrt{\frac{\zeta}{L_1}}>\zeta$ \eqref{AdiC}\\
\hline $\zeta\geq x_{+}$ & $L_1=0$ & $L_1=x_{-}$\\
\cline{2-3} (barrier \emph{ineffective}) & $A=\zeta=4$ & $A=\zeta=x_{+}$\\
\hline
\end{tabular}
\caption{Values of $L_1$ and $A$ in the different regions of the
$(c,\zeta)$ plane.}
\end{table}

The support of $\hf$ is:
\begin{equation}\label{Support}
  L_1(c,\zeta)\leq x\leq \min[\zeta, A(c,\zeta)]
\end{equation}

At the lower edge of the support $L_1(c,\zeta)$, the density
vanishes \emph{unless} $c=1$, in which case it diverges as $\sim
1/\sqrt{x}$.

At the upper edge of the support, according to the value of the
minimum ($\zeta$ or $A(c,\zeta)$) in \eqref{Support} the density
can respectively diverge as $\sim 1/\sqrt{\zeta-x}$ or vanish.

Note that the unconstrained Mar\v{c}enko-Pastur law
\eqref{Marcenko-Pastur} is recovered from \eqref{Generic form}
when the barrier is ineffective, i.e. $\zeta\geq
x_{+}$.
%$L_1(c,\zeta)= x_{-}$ and $A(c,\zeta)=\zeta=x_{+}$. We will
%show later that this indeed happens when the barrier $\zeta\to
%x_{+}(c)=\left(\frac{1}{\sqrt c}+1\right)^2$ from below.

\subsection{The $c=1$ case}\label{Subsection IIIa}
In this case, the support of the unconstrained spectral density is
$(0,4]$ and the Mar\v{c}enko-Pastur law prescribes an inverse
square root divergence at $x=0$ . Furthermore, the density
vanishes at $x=4$ (see Fig. (\ref{fig:mp})).

In the constrained case, the barrier at $\zeta$ is only effective
when $0\leq\zeta\leq 4$. When the barrier crosses the point
$\zeta=4$ from below, the density shifts back again to the
unconstrained case.

\begin{figure}[htb]
\includegraphics[bb=91 3 322 146,totalheight=0.2\textheight]{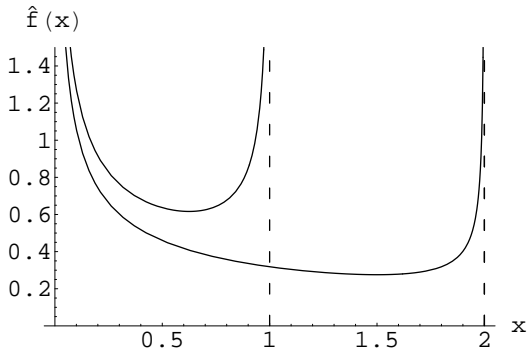}
\caption{Constrained spectral density $\hf(x)$ for the barrier at
$\zeta=1$ and $\zeta=2$. \label{fdix}}
\end{figure}

The integral equation for $\hf(x)$ \eqref{Stationarity3} becomes:
\begin{equation}\label{IntegralEqc=1}
  \frac{1}{2}=\mathcal{P}\int_0^\zeta\frac{\hf(x^\prime)}{x-x^\prime}dx^\prime\qquad 0\leq x\leq
  \zeta
\end{equation}
The general solution of equations of the type:
\begin{equation}\label{IntegralEqc=2}
  \mathcal{P}\int_0^\zeta\frac{\hf(x^\prime)}{x-x^\prime}dx^\prime = g(x)
\end{equation}
is given by Tricomi's theorem~\cite{Tricomi1}:
\begin{equation}\label{Tricomi}
  \hf(x)=\frac{1}{\pi^2\sqrt{x(\zeta-x)}}\left[\mathcal{P}\int_0^\zeta \sqrt{\omega(\zeta-\omega)}\frac{g(\omega)}{\omega-x}d\omega + B\right]
\end{equation}
where $B$ is an arbitrary constant. After putting
$g(\omega)=1/2$ in \eqref{Tricomi} and determining $B$ by the
normalization condition
$\int_0^\zeta \hf(x)dx=1$ we finally get:
\begin{equation}\label{Constrained density}
  \hf(x)=\frac{1}{2\pi\sqrt{x(\zeta-x)}}\left[\frac{\zeta}{2}+2-x\right]\qquad 0\leq x\leq
  \zeta
\end{equation}
A plot of this charge density for two values of the barrier
$\zeta$ is given in Fig. \ref{fdix}.
In summary, the average density of states with a barrier at
$\zeta$ is given by:
\begin{align}
\hf(x)=
\begin{cases}
\frac{1}{2\pi\sqrt{x(\zeta-x)}}\left[\frac{\zeta}{2}+2-x\right] &
0\leq\zeta\leq 4\\ \label{constr}
  \frac{1}{2\pi}\sqrt{\frac{4-x}{x}} & \zeta\geq 4
\end{cases}
\end{align}
Thus, for all $\zeta>4$, the solution sticks to the $\zeta=4$
case. Note that both cases in \eqref{constr} can be obtained from
the general formula \eqref{Generic form}.

\begin{figure}[htb]
\includegraphics[bb=91.5625 3.1875 321.938 164.438,totalheight=0.2\textheight]{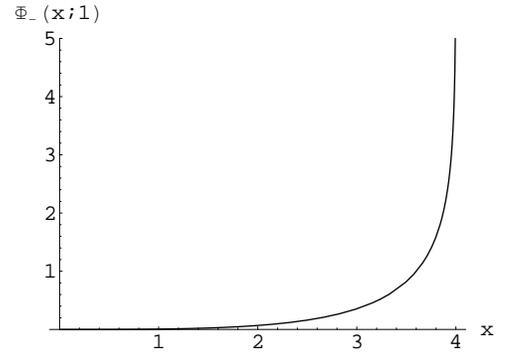}
\caption{Rate function $\Phi_{-}(x;1)$ (see \eqref{Large deviation
function}).\label{Phi}}
\end{figure}

Now we can substitute \eqref{constr} back into
\eqref{Stationarity2} to find the value of the multiplier $C_1$
and eventually evaluate the action $S[f^\star(x);\zeta]$ \eqref{S}
explicitly for $0\leq\zeta\leq 4$:
\begin{equation}\label{ActionStar}
    S(\zeta):=S[\hf^\star(x);\zeta]=2\log 2-\log \zeta+\frac{\zeta}{2}-\frac{\zeta^2}{32}
\end{equation}
From \eqref{Steep}, we get $Z_1(\zeta)\approx\exp(-\beta N^2
S(\zeta)/2)$. For the denominator,
$Z_0=Z_1(\zeta=\infty)=Z_1(\zeta=4)\approx \exp(-\beta N^2
S(4)/2)$, where we have used the fact that the solution for any
$\zeta>4$ (e.g., when $\zeta=\infty$) is the same as the solution
for $\zeta=4$. Thus, eventually the probability $P_N(t)$
\eqref{P_N(t)} decays for large $N$ as:
\begin{align}\label{Scaling of P_N(t)}
  \nonumber P_N(t) &=\frac{Z_1(t)}{Z_0}\approx
  \exp\left\{-\frac{\beta}{2}N^2[S(\zeta)-S(4)]\right\}\\
  &\approx\exp\left\{-\frac{\beta}{2}N^2\Phi_{-}\left(\frac{4N-t}{N};1\right)\right\}
\end{align}
where the rate function is given by
\begin{equation}\label{Large deviation function}
  \Phi_{-}(x;1)=
  \begin{cases}
  2\log 2-\log(4-x)-\frac{x}{4}-\frac{x^2}{32} & x\geq 0\\
  0 & x\leq 0
  \end{cases}
\end{equation}
and is plotted in fig.\ref{Phi}

We now turn to the original problem of determining the probability
of the following extremely rare event, i.e. that all the
eigenvalues happen to lie below the mean value
$\langle\lambda\rangle=\int_0^{4N}\lambda\rho_N(\lambda)d\lambda=N$.
Starting from \eqref{Scaling of P_N(t)}, this is easily computed
by putting the barrier at the mean value $t=N$, i.e., $\zeta=1$.
We then get for large $N$
\begin{equation}
{\rm Prob}\left[\lambda_{\rm max}\le \langle \lambda\rangle=N,
N\right]\sim \exp[-\theta(1) N^2] \label{theta11}
\end{equation}
where
\begin{eqnarray}
\theta(1)&=& \frac{\beta}{2} \Phi_{-}\left(3;1\right) \nonumber \\
&=& \beta\,\left(\log 2-\frac{33}{64}\right). \label{theta12}
\end{eqnarray}

Since we are calculating here the probability of negative
fluctuations of $\lambda_{\rm max}$ of $O(N)$ to the left of the
mean $\langle \lambda_{\rm max}\rangle=x_{+}(c) N$, when the
argument of the left rate function $\Phi_{-}(x;1)$ is very small
(i.e., for fluctuations $\ll O(N)$), \eqref{Scaling of P_N(t)}
should smoothly match the left tail of the Tracy-Widom
distribution that describes fluctuations of order $\sim
O(N^{1/3})$ to the left of the mean $\langle \lambda_{\rm
max}\rangle=x_{+}(c) N$. Indeed, from \eqref{Large deviation
function} as $x\to 0$
\begin{equation}
\Phi_{-}(x;1) \approx \frac{x^3}{192} \label{smallphiy}
\end{equation}
and substituting \eqref{smallphiy} in \eqref{Scaling of P_N(t)} we
get, for fluctuations $\ll O(N)$ to the left of the mean,
\begin{eqnarray}
P_N(t)&\sim &\exp\left[-\frac{\beta}{384}\,N^2\, (4-t/N)^3\right] \nonumber \\
&=& \exp\left[-|\chi|^3/12\right] \label{ctw1}
\end{eqnarray}
where $\chi= (t-4N)/(2^{4/3} N^{1/3})$. This coincides with
Johansson's result for the Tracy-Widom fluctuations in \eqref{JJ}
for $c=1$ and comparing \eqref{ctw1} and \eqref{asymp1}, we see
that we recover the left tail of the Tracy-Widom distribution.

\subsection{The $c<1$ case}\label{Subsection IIIb}
Our approach is very similar to the previous case. However, some
additional technical subtleties, which we emphasize, arise in this
case.

As in the unconstrained case, we expect a lower bound $L_1\equiv
L_1(c,\zeta)$ to the support of the constrained $\hf(x)$. The
parameter $L_1$ will be determined later through the normalization
condition for $\hf(x)$.

It is convenient to reformulate \eqref{Stationarity3} in terms of
the new variable $y=x-L_1$, measuring the distance with respect to
the lower edge of the support, where $\hf(x)$ is assumed to
vanish.

Equation \eqref{Stationarity3} then reads:
\begin{equation}\label{Station}
  \frac{1}{2}-\frac{\alpha}{2(y+L_1)}=\mathcal{P}\int_0^L\frac{\tilde{f}(y^\prime)}{y-y^\prime}dx^\prime\qquad 0\leq y\leq
  L
\end{equation}
where we have denoted $L=\zeta-L_1$ and $\tilde{f}(y)\equiv
\hf(y+L_1)$.

The general solution of \eqref{Station} in this case is:
\begin{equation}\label{ftilde}
  \tilde{f}(y)=\frac{1}{\pi\sqrt{y(L-y)}}\left[-\frac{y}{2}-\frac{\alpha}{2}\frac{\sqrt{L_1(L+L_1)}}{y+L_1}+B^\prime\right]
\end{equation}
and the constant $B^\prime$ is determined by the condition
$\tilde{f}(y=0)=0$. Thus we get:
\begin{equation}\label{ftildefinal}
    \tilde{f}(y)=\frac{\sqrt{y}}{2\pi\sqrt{L-y}}\left[\frac{A-L_1-y}{y+L_1}\right]
\end{equation}
where:
\begin{equation}\label{AdiC}
A\equiv A(c,\zeta)=\alpha\sqrt{\zeta/L_1}
\end{equation}

Note that everything is expressed in terms of the only still
unknown parameter $L_1$.

From \eqref{ftildefinal} we can infer two kinds of possible
behaviors for $\tilde{f}(y)$ due to the competing effects of the
singularity for $y\rightarrow L$ (where the denominator vanishes)
and the suppression for $y\rightarrow A-L_1$ (where the numerator
vanishes).

Thus, depending on which of the following two conditions applies
once we have put the barrier at $\zeta$:
\begin{align}\label{conditions}
\nonumber A(c,\zeta)-L_1(c,\zeta)>L(c,\zeta) &\rightarrow
\sqrt{L_1(c,\zeta)\zeta}<\alpha &\quad\mathrm{(I)}\\
A(c,\zeta)-L_1(c,\zeta)<L(c,\zeta) &\rightarrow
\sqrt{L_1(c,\zeta)\zeta}>\alpha &\quad\mathrm{(II)}
\end{align}
$\tilde{f}$ can diverge at $y=L$ or vanish at $A-L_1$
respectively. In \eqref{conditions} we have restored the
functional dependence for clarity.

This is a subtle point because, given the barrier at $\zeta$, we
cannot determine \emph{a priori} which of the previous conditions
holds. In fact, $L_1(c,\zeta)$ should be determined \emph{a
posteriori} separately for each case from the normalization
condition:
\begin{equation}\label{NormCond}
\int_0^L \tilde{f}(y)dy=1
\end{equation}
Once this is done, it turns out that the conditions
\eqref{conditions} can be reformulated in terms of the position of
the barrier $\zeta$ in the following much simpler way:
\begin{align}\label{conditions2}
\nonumber 0<\zeta<x_{+}&\quad\mathrm{(I)}\\
\zeta\geq x_{+}&\quad\mathrm{(II)}
\end{align}

We summarize here the final results in the two cases.
\subsubsection{Case I.   $0<\zeta<x_{+}$}\label{I}
The normalization condition \eqref{NormCond} leads to the
following cubic equation for $w\equiv
w(c,\zeta)=\sqrt{L_1(c,\zeta)}$:
\begin{equation}\label{Cubic Equation}
  w^3-[2(2+\alpha)+\zeta]w+2\alpha\sqrt{\zeta}=0
\end{equation}
which has always three real solutions, one negative ($w_0$) and
two positive:
\begin{equation}\label{SolutionsCubic}
  w_k(c,\zeta)=\frac{2p}{3\varrho^{1/3}}\cos\left(\frac{\theta+2k\pi}{3}\right)\qquad
  k=0,1,2
\end{equation}
where:
\begin{equation}\label{Parameters}
\begin{cases}
\nonumber p &=-[2(2+\alpha)+\zeta]\\
\nonumber q &=2\alpha\sqrt{\zeta}\\
\nonumber B &=-\left(\frac{q^2}{4}+\frac{p^3}{27}\right)\\
\nonumber \varrho &=\sqrt{-p^3/27}\\
\theta &=\arctan\left(\frac{2\sqrt{B}}{q}\right)
\end{cases}
\end{equation}
Note that $w_2<w_1$. With simple considerations, we conclude that
the right root to be chosen is $w_2(c,\zeta)$. Thus:
\begin{equation}\label{L1}
  L_1(c,\zeta)=w_2^2(c,\zeta)
\end{equation}

Finally, we can write down the full constrained unshifted spectral
density as:
\begin{equation}\label{FullConstrainedSpectralDensity}
  \hf(x)=\frac{1}{2\pi}
\frac{\sqrt{x-L_1(c,\zeta)}}
{\sqrt{\zeta-x}}\left[\frac{A(c,\zeta)-x}{x}\right]
\end{equation}
valid for $L_1(c,\zeta)\leq x\leq \zeta$ where $L_1(c,\zeta)$ is
given by \eqref{L1} and $A(c,\zeta)$ by \eqref{AdiC}.

\begin{figure}[htb]
\includegraphics[bb= 91 3 322 146,totalheight=0.23\textheight]{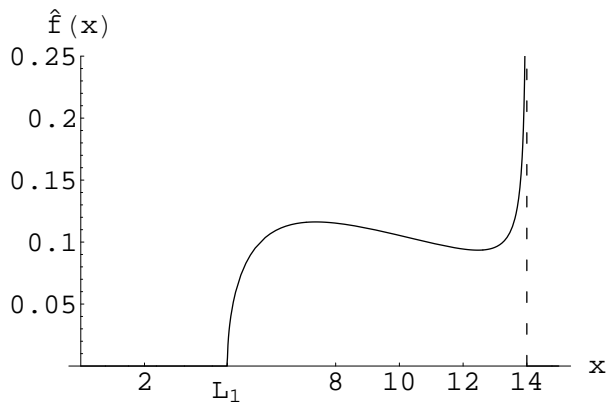}
\caption{Constrained spectral density $\hf(x)$ for $c=0.1$ and
$\zeta=14$.\label{effex}}
\end{figure}

A plot of $\hf(x)$ for $c=0.1$ and $\zeta=14$ is given in fig.
\ref{effex}. In this case, $L_1(c,\zeta)\approx 4.60084$ and
$A(c,\zeta)\approx 15.6996$.
\subsubsection{Case II.   $\zeta\geq x_{+}$}\label{II}
In this case, the barrier is immaterial and we should recover the
unconstrained Mar\v{c}enko-Pastur distribution.
The support of
$\tilde{f}(y)$ is $[0, A-L_1]$
and this implies that we can safely put $L=A-L_1$ in \eqref{NormCond}.

The integration can be performed and coming back to the unshifted
spectral density $\hf(x)$ we get:
\begin{equation}\label{fdixbarrier}
  \hf(x)=\frac{1}{2\pi}\frac{\sqrt{x-L_1}\sqrt{L_2-x}}{x}
\end{equation}
valid for $L_1\leq x\leq L_2$ where:
\begin{equation}\label{L1andL2}
  \begin{cases}
  L_1 & = x_{-}\\
  L_2 & = L_1+L =x_{+}
  \end{cases}
\end{equation}
which is the unconstrained Mar\v{c}enko-Pastur distribution, as it
should.\\
\\
\\
It is interesting to evaluate the limit $c\rightarrow 1^{-}$ in
\eqref{FullConstrainedSpectralDensity} and \eqref{fdixbarrier} in
order to recover the result \eqref{constr} in subsection
\ref{Subsection IIIa}. The case of equation \eqref{fdixbarrier} is
obvious. For the other, it is a matter of simple algebra to show
that:
\begin{align}\label{Limit}
  \lim_{c\rightarrow 1^{-}}L_1(c,\zeta) &=0\\
  \lim_{c\rightarrow 1^{-}}A(c,\zeta) &=(\zeta+4)/2
\end{align}
so that \eqref{FullConstrainedSpectralDensity} matches
\eqref{constr}.

Furthermore, Cases I and II should match smoothly as
$\zeta$ hits precisely the limiting value $x_{+}$. This
corresponds to $A(c,\zeta)\equiv\zeta\rightarrow A(c,x_{+})\equiv
x_{+}$. It is again straightforward to check that this last
condition implies $L_1(c,x_{+})\equiv x_{-}$ so that
\eqref{FullConstrainedSpectralDensity} recovers
\eqref{fdixbarrier}.

\begin{figure}[htb]
\includegraphics[bb=91.5625 3.1875 321.938 233.563,totalheight=0.27\textheight]{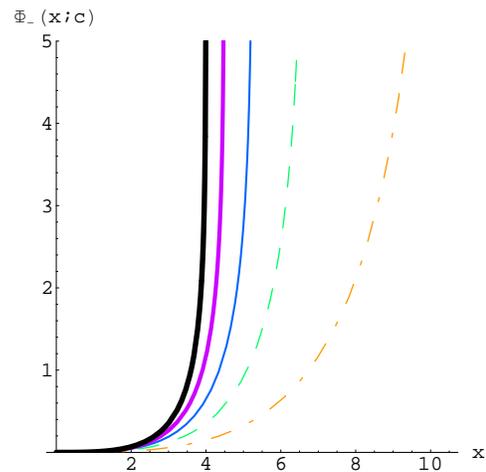}
\caption{Rate function $\Phi_{-}(x;c)$ for the following values
(from left to right) of $c=1,0.8,0.6,0.4,0.2$. See also Figure
\ref{Phi}.\label{Phicmin1}}
\end{figure}

The interesting case for computing large fluctuations is Case I.
One can insert \eqref{FullConstrainedSpectralDensity} into
\eqref{S} in order to evaluate \eqref{Steep}. It turns out that
the integrals involved can be analytically solved in terms of
derivatives of hypergeometric functions, but a more explicit
formula is derived in the Appendix. We give here a plot of the
rate function $\Phi_{-}(x;c)$ that describes the large
fluctuations of $O(N)$ to the left of the mean $\langle
\lambda_{\rm max}\rangle = x_{+}(c) N$:
\begin{align}\label{large deviation function2}
  \nonumber P_N(t)
  &=\frac{Z_1(t)}{Z_0}\approx\exp\left\{-\frac{\beta}{2}N^2[S(\zeta)-S(x_{+})]\right\}\\
  &=\exp\left\{-\frac{\beta}{2}N^2\Phi_{-}\left(x_{+}-\frac{t}{N};c\right)\right\}
\end{align}
The plot is given in Fig. \ref{Phicmin1} for several values of $c$
approaching $1$. The limiting case $\Phi_{-}(x;1)$ \eqref{Large
deviation function} is also plotted.

We can now compute to leading order the probability that all the
eigenvalues are less than the mean value
$\langle\lambda\rangle=N/c$. This amounts to putting the barrier
at $t=N/c$ in \eqref{large deviation function2}, which gives
$\Phi_{-}\left(\frac{2}{\sqrt c}+1;c\right)$. Several numerical
values are given in the following table.
\begin{table}[htb]
\begin{tabular}{|c|c|}
\hline
  $c$  & $\Phi_{-}\left(\frac{2}{\sqrt{c}}+1;c\right)$ \\ \hline
  0.1 & 0.475802 \\
  0.2 & 0.449162 \\
  0.4 & 0.414592 \\
  0.6 & 0.390245 \\
  0.8 & 0.37104 \\
  0.95 & 0.358805 \\ \hline
  1 & 0.355044 \\ \hline
\end{tabular}
\caption{Some values of the rate function (see text for further
explanation).}
\end{table}

\section{Numerical Results}\label{Section IV}
The formulas \eqref{constr},\eqref{Scaling of P_N(t)},
\eqref{FullConstrainedSpectralDensity} and \eqref{large deviation
function2} have been numerically checked on samples of hermitian
matrices ($\beta=2$) up to $N=30$, $M=300$ and the agreement with
the analytical results is already excellent. We describe in this
section the numerical methods and results.

A direct sampling of Wishart matrices up to those sizes is
computationally very demanding. We applied the following much
faster technique, suggested in \cite{DE}.

Let $L_\beta=B_\beta B^T_\beta$ be the tridiagonal matrix
corresponding to:
\begin{equation}
B_\beta\sim
\begin{pmatrix}
   \chi_{2a}& \ & \ & \ \\
  \chi_{\beta(N-1)} & \chi_{2a-\beta} & \  & \  \\
  \  & \ddots & \ddots & \  \\
  \  & \ & \chi_{\beta} & \chi_{2a-\beta(N-1)}
\end{pmatrix}
\end{equation}
$B_\beta$ is a square $N\times N$ matrix with nonzero entries on
the diagonal and subdiagonal and $a=(\beta/2)M$. The nonzero
entries $\chi_k$ are independent random variables obtained from
the square root of a $\chi^2$-distributed variable with $k$
degrees of freedom. It has been proved in \cite{DE} that $L_\beta$
has the same joint probability distribution of eigenvalues as
\eqref{jpdWishart}. Thus, as far as we are interested in
eigenvalue properties, we can use the $L_\beta$ ensemble instead
of the original Wishart one. This makes the diagonalization
process much faster due to the tridiagonal structure of the
matrices $L_\beta$.

\begin{figure}[htb]
\includegraphics[bb = 52 198 549 589,totalheight=0.3\textheight]{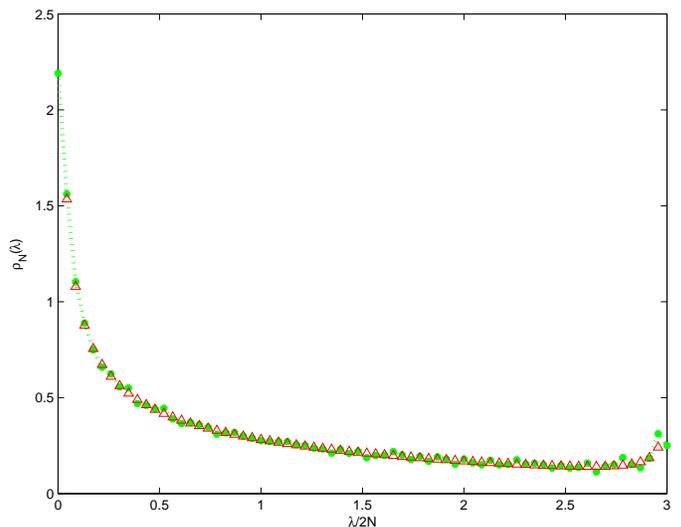}
\caption{Constrained spectral density $\hat\varrho_N(\lambda)$ for
$N=M=30$. The barrier is at $\zeta=3$. In dotted green the
histogram of rescaled eigenvalues over an initial sample of
$3\times 10^5$ matrices ($\beta=2$). In triangled red the
theoretical distribution. \label{N=M=30}}
\end{figure}

\begin{figure}[htb]
\includegraphics[bb=23 187 549 615,totalheight=0.3\textheight]{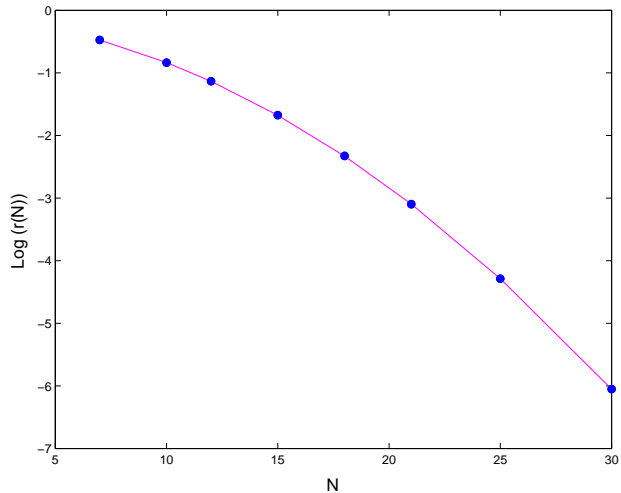}
\caption{Natural logarithm of the probability that all the
rescaled eigenvalues are less than $\zeta=3$ vs. $N$ for the case
$c=1$ ($x_{+}=4$). The data points are fitted with a parabola
(solid line).\label{parabola c=1}}
\end{figure}

\begin{figure}[htb]
\includegraphics[bb=40 194 549 589,totalheight=0.27\textheight]{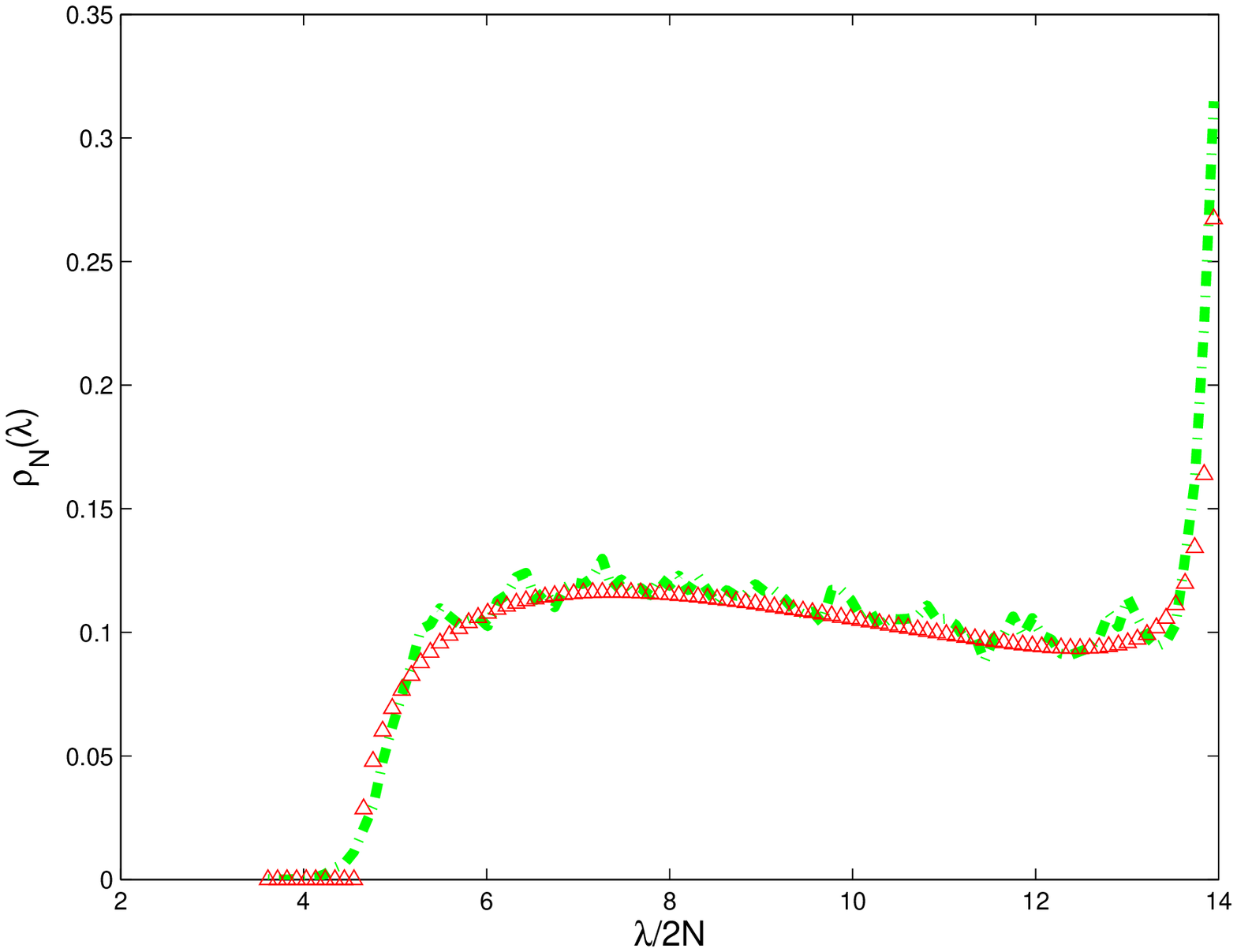}
\caption{Constrained spectral density $\hat\varrho_N(\lambda)$ for
$N=10$, $M=100$ ($c=0.1$). The barrier is at $\zeta=14$. In
dash-dotted green the histogram of rescaled eigenvalues over an
initial sample of $5\times 10^5$ matrices ($\beta=2$). In
triangled red the theoretical distribution.\label{N=10 M=100}}
\end{figure}

\begin{figure}[htb]
\includegraphics[bb=55 195 549 589,totalheight=0.27\textheight]{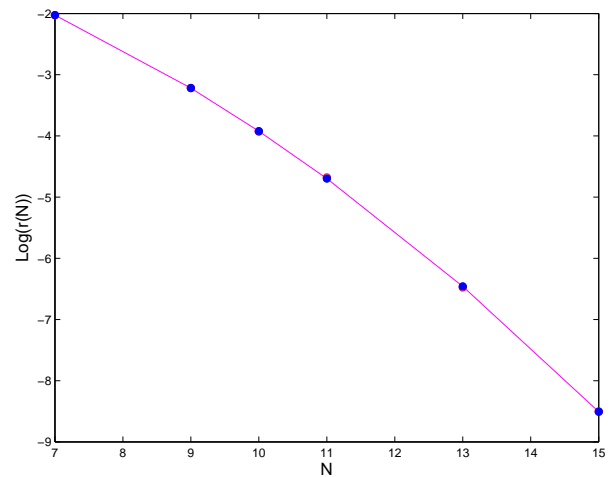}
\caption{Natural logarithm of the probability that all the
rescaled eigenvalues are less than $\zeta=14$ vs. $N$ for the case
$c=0.1$ ($x_{+}\approx 17.32$). The data points are fitted with a
parabola (solid line).\label{parabola clessthan1}}
\end{figure}

We report the following four plots: the first two (fig.
\ref{N=M=30} and \ref{parabola c=1}) are for the case $c=1$ and
the last two (fig. \ref{N=10 M=100} and \ref{parabola clessthan1})
for the case $c=0.1$.

In fig. \ref{N=M=30}, we plot the histogram of normalized
eigenvalues $\lambda/2N$ for an initial sample of $3\times 10^5$
hermitian matrices ($\beta=2$, $N=M=30$), such that matrices with
$\lambda_{\rm max}/2N>\zeta$ are discarded. The barrier is located
at $\zeta=3$. On top of it we plot the theoretical distribution
\eqref{constr}.

In fig. \ref{N=10 M=100}, we do the same but in the case $N=10,
M=100$. The barrier is located at $\zeta=14$. The theoretical
distribution is now taken from
\eqref{FullConstrainedSpectralDensity}.

To obtain the plots in fig. \ref{parabola c=1} and \ref{parabola
clessthan1}, we generate $\approx 5\times 10^5$ $L_2$ matrices for
different values of $N=7\rightarrow 30$ (or $15$). The parameters
$(c,\zeta)$ are kept fixed to the value $(1,3)$ for fig.
\ref{parabola c=1} ($x_{+}=4$) and $(0.1,14)$ for fig.
\ref{parabola clessthan1} ($x_{+}\approx 17.32$). The constraining
capability of those barriers can be estimated by the ratio
$\kappa(c,\zeta)=(x_{+}-\zeta)/(x_{+}-x_{-})$, corresponding to
the window of forbidden values for the largest eigenvalue. We get
$\kappa(1,3)=0.25$ and $\kappa(0.1,14)\approx 0.26$, to be
compared with the values of $\kappa(c,\zeta)=(2+\sqrt{c})/4$ for
the barrier at the mean value $\zeta=1/c$, which would give
respectively $\kappa=0.75$ and $\kappa\approx 0.58$. This relative
mildness of the constraint allows us to get a much more reliable
and faster statistics in the simulations.

For each value of $N$, we determine the empirical frequency $r(N)$
of constrained matrices as the ratio between the number of
matrices whose largest rescaled eigenvalue is less than $\zeta$
and the total number of samples $(5\times 10^5)$. The logarithm of
$r(N)$ vs. the size $N$ is then naturally fitted by a parabola $a
N^2+b N+\hat{c}$ to test the prediction for $a$ in formulas
\eqref{Scaling of P_N(t)} and \eqref{large deviation function2}.

The best values for the coefficient $a$ of the leading term are
estimated as $-0.006153$ ($c=1$) and $-0.0357$ ($c=0.1$), to be
compared respectively with the theoretical prediction
$\Phi_{-}(1;1)\approx -0.006432$ and
$\Phi_{-}(x_{+}-14;0.1)\approx -0.03666$. Despite the relatively
small sizes and the $O(N)$ corrections, the agreement is already
good.


\section{Conclusions}\label{Section V}

In this paper we have studied the probability of atypically large
negative fluctuations (with respect to the mean) of the largest
eigenvalue $\lambda_{\rm max}$ of a random Wishart matrix. The
standard Coulomb gas analogy for the joint probability
distribution of eigenvalues allowed us to use the tools of
statistical physics, such as the functional integral method
evaluated for large $N$ by the method of steepest descent. Using
these tools, we have analytically computed the probability of
large deviations of $\lambda_{\rm max}$ to the left of its mean.
In particular, our main motivation was to compute the probability
of a rare event: all eigenvalues are less than the average
$\langle \lambda\rangle=N/c$. This implies that the largest
eigenvalue itself is less than $\langle \lambda\rangle=N/c$. This
question is relevant in estimating the efficiency of the
`principal components analysis' method used in multivariate
statistical analysis of data. Our main result is to show that, to
leading order in $N$, this probability decays as
$\sim\exp[-\frac{\beta}{2}N^2 \Phi_{-}(\frac{2}{\sqrt{c}}+1;c)]$,
where $\Phi_{-}(x;c)$ is a rate function that we have explicitly
computed. The quadratic, instead of linear, $N$-dependence of the
exponential reflects the eigenvalue correlations.

Furthermore, our method allows us to determine exactly the
functional form of the constrained spectral density, i.e., the
average charge density of a Coulomb gas constrained to be within a
finite box $\lambda\in [0,t]$.

All the analytical results are in excellent agreement with the
numerical simulations on samples of hermitian matrices up to
$N=30$, and the estimates of the large deviation prefactor are
already good even for $N\sim 15$. \vspace{10pt}
\begin{acknowledgments}
One of us (PV) has been supported by a Marie Curie Early Stage
Training Fellowship (NET-ACE project). We are grateful to Gernot
Akemann, Igor Krasovsky and Yang Chen for helpful comments and for
pointing to us relevant references. The support by Sergio Consoli
(Brunel) for the numerical simulations is also gratefully
acknowledged.
\end{acknowledgments}

\appendix*
\section{Rate function for $c<1$}
We evaluate in closed form the action
$S(\zeta):=S[\hf^\star(x);\zeta]$ (see \eqref{S}) for the case
$c<1$, where $\hf^\star(x)$ is given by
\eqref{FullConstrainedSpectralDensity}. The result is in eq.
\eqref{FinalS}.

The rate function $\Phi_{-}(x; c)$ for $c<1$, given by:
\begin{equation}\label{RateFunctionC<1}
  \Phi_{-}(x;c)=S\left(x_{+}-x\right)-S\left(x_{+}\right)
\end{equation}
can be evaluated immediately.

After inserting \eqref{FullConstrainedSpectralDensity} into
\eqref{S} and determining $C_1$ from \eqref{Stationarity2}, we
find that $S(\zeta)$ is given by:
\begin{align}\label{Sfinalclessthan1app}
  \nonumber S(\zeta) &=\frac{1}{2}\int_{L_1}^\zeta \hf(x) x
  dx-\frac{\alpha}{2}\int_{L_1}^\zeta \hf(x) \log(x)
  dx+\\
  &-\int_{L_1}^\zeta \hf(x) \log(x-L_1)
  dx+\frac{L_1}{2}-\frac{\alpha}{2}\log(L_1)
\end{align}
After the substitution $x=(\zeta-L_1)t+L_1$ in the integrals in
\eqref{Sfinalclessthan1app} and some simple algebra, $S(\zeta)$
can be expressed as:
\begin{equation}\label{FinalS}
  S(\zeta)=-\frac{\alpha}{2}\Theta_1-\Theta_2+\frac{\zeta-L_1}{2}\Xi+\frac{L_1}{2}-\frac{\alpha}{2}\log(L_1)
\end{equation}
where $\Theta_k$ and $\Xi$ are the following functions of $c$ and
$\zeta$:
\begin{align}\label{ThetaXi}
  \nonumber\Theta_k &= \frac{\zeta-L_1}{2\pi}\left\{\log(\zeta-L_1)
\left[\frac{A}{\zeta-L_1}\mathcal{I}_0\left(\frac{L_1}{\zeta-L_1}\right)-\frac{\pi}{2}\right]
  +\right. \\
  \nonumber &+\frac{A}{\zeta-L_1}\mathcal{I}_k\left(\frac{L_1}{\zeta-L_1}\right)\left. \right\} \\
  \Xi &= \frac{A}{4}-\frac{3}{16}\zeta-\frac{L_1}{16}+\frac{\alpha}{2\pi}\mathcal{I}_3\left(\frac{L_1}{\zeta-L_1}\right)
  +\frac{1}{2}-\log(2)
\end{align}
The functions $\mathcal{I}_k(x)$ are given by the following
integrals:
\begin{align}
  \mathcal{I}_0(x) &= \frac{d}{dx}\mathcal{I}_3(x) \\
  \mathcal{I}_1(x) &= \int_0^1 dt
  \frac{\log(t+x)}{t+x}\sqrt{\frac{t}{1-t}} \label{Into1}\\
  \mathcal{I}_2(x) &= \int_0^1 dt
  \frac{\log t}{t+x}\sqrt{\frac{t}{1-t}}\label{Into2}\\
  \mathcal{I}_3(x) &= \int_0^1 dt
  \log(t+x)\sqrt{\frac{t}{1-t}}\label{Into3}
\end{align}
which, following very closely ref. \cite{Chen} paper I, appendix
B, can be computed explicitly in closed form.

The integral $\mathcal{I}_3(x)$ (and thus also $\mathcal{I}_0(x)$)
can be computed by Mathematica$^{\circledR}$:
\begin{align}\label{I3}
  \nonumber \mathcal{I}_3(x) &= \frac{\pi}{2}\left[1+2x-2\sqrt{x(1+x)}+2\log\left[1+\sqrt{1+\frac{1}{x}}\right]+\right.\\
  &+\log\left(\frac{x}{4}\right)\left.\right]
\end{align}
while $\mathcal{I}_1(x)$ and $\mathcal{I}_2(x)$ are given in terms
of derivatives of hypergeometric functions. More explicit
expressions can be given as follows, starting with
$\mathcal{I}_1(x)$. Exploiting the identity $h^\lambda \log h =
\partial_\lambda h^\lambda$, we can rewrite the integral as:
\begin{equation}\label{IntegralRewritten}
\mathcal{I}_1(x)=\left[\partial_\lambda\int_0^1 dt
(t+x)^\lambda\sqrt{\frac{t}{1-t}}\right]\Big|_{\lambda=-1}
\end{equation}
and the integral in \eqref{IntegralRewritten} can be evaluated in
terms of Kummer's hypergeometric function:
\begin{equation}\label{Kummer}
  \mathcal{I}_1(x)=\frac{\pi}{2}\left\{\partial_\lambda\left[x^\lambda~_2 F_1\left(-\lambda,\frac{3}{2};2;-\frac{1}{x}\right)\right]\right\}\Big|_{\lambda=-1}
\end{equation}
Now, applying the transformation formulas \cite{Abramo} [15.3.7
pag. 559] and evaluating the derivatives of Gamma functions that
arise, we finally get:
\begin{align}\label{IntegralRewrittenAfterTransform}
\nonumber\mathcal{I}_1(x) &=\frac{\pi}{2}\left[-2\log
4+2~\hat{i}_1(x)-2\sqrt{\frac{x}{1+x}}\log\left(\frac{4x}{e^2}\right)+\right.\\
&-2\sqrt{x}~\hat{i}_2(x)\left.\right]
\end{align}
where:
\begin{align}\label{i1and2}
  \hat{i}_1(x) &= \left[\partial_\mu~_2 F_1(1-\mu,-\mu;-\mu+1/2;-x)\right]\Big |_{\mu=0} \\
\hat{i}_2(x) &= \left[\partial_\mu~(1+x)^{\mu-1/2}~_2
F_1(\mu,\mu+1;\mu+3/2;-x)\right]\Big |_{\mu=0}
\end{align}
To evaluate $\hat{i}_1(x)$ and $\hat{i}_2(x)$, we use the series
expansion for hypergeometric functions \cite{Abramo} [15.1.1 pag.
556] and upon differentiation we get:
\begin{equation}\label{i1}
  \hat{i}_1(x)=-\sum_{n=1}^\infty B\left(\frac{1}{2},n
  \right)(-x)^n
\end{equation}
where $B(v,w)=\frac{\Gamma(v)\Gamma(w)}{\Gamma(v+w)}$ is Euler's
Beta function. Introducing the integral representation of the Beta
function:
\begin{equation}\label{IntegralBeta}
  B(x,y)=\int_0^1 dt~ t^{x-1}(1-t)^{y-1}
\end{equation}
into \eqref{i1} and upon exchanging summation and integral, we
arrive with the help of $\sum_{n=0}^\infty (-xt)^n=(1+xt)^{-1}$
to:
\begin{equation}\label{i1new}
  \hat{i}_1(x)=x\int_0^1
  \frac{dt}{\sqrt{1-t}(1+xt)}=2\sqrt{\frac{x}{1+x}}\mbox{arcsinh}(\sqrt{x})
\end{equation}
Following the same procedure, we get for $\hat{i}_2(x)$:

\begin{equation}\label{i1new}
  \hat{i}_2(x)=\frac{1}{\sqrt{1+x}}\left[\log(1+x)-i_1(x)\right]
\end{equation}
where $i_1(x)$ is defined in ref. \cite{Chen} as:
\begin{equation}\label{i1chen}
  i_1(x)=-2+2\sqrt{\frac{1+x}{x}}\mbox{arctanh}\left(\sqrt{\frac{x}{1+x}}\right)
\end{equation}
From \eqref{IntegralRewrittenAfterTransform} we get the final
result for $\mathcal{I}_1(x)$:
\begin{align}\label{FinalI1}
  \nonumber\mathcal{I}_1(x) &=\pi\left\{
  -\log 4 +\sqrt{\frac{x}{1+x}}\left[
  2\mbox{arcsinh}(\sqrt{x})+\right.\right.\\
  &+2\sqrt{1+\frac{1}{x}}\mbox{arctanh}\left(\sqrt{\frac{x}{1+x}}\right)
  -\log[4x(1+x)]\left.\left.\right]\right\}
\end{align}

Following the very same procedure as in the previous case, we find
for $\mathcal{I}_2(x)$:
\begin{equation}\label{FinalI2}
  \mathcal{I}_2(x)=\pi\left[-\log
  4+\sqrt{\frac{x}{x+1}}\left(2\mbox{arcsinh}(\sqrt{x})-\log(x)\right)\right]
\end{equation}
Now we compute the limit $c\rightarrow 1^{-}$ in \eqref{FinalS} to
recover \eqref{ActionStar}. Given that, for $c\rightarrow 1^{-}$,
$L_1\rightarrow 0$, $\alpha\rightarrow 0$ and $A\rightarrow
(\zeta+4)/2$, we have to evaluate the integrals $\mathcal{I}_k(x)$
for $x\rightarrow 0$. This gives:
\begin{align}\label{limiting integral}
  \mathcal{I}_0(0) &\sim  \pi\\
  \mathcal{I}_1(0) &\sim  -\pi\log 4\\
  \mathcal{I}_2(0) &\sim  -\pi\log 4\\
  \mathcal{I}_3(0) &\sim  -\frac{\pi}{2}(\log 4 -1)
\end{align}
Then, $S[\hf^\star(x);\zeta]\Big|_{c\rightarrow 1^{-}}\sim
\left[-\Theta_2+\frac{\zeta}{2}\Xi\right]\Big|_{c\rightarrow
1^{-}}=2\log 2 -\log\zeta+\frac{\zeta}{2}-\frac{\zeta^2}{32}$ as
it should (see \eqref{ActionStar}).


\vspace{10pt}






































\begin{thebibliography}{99}
\bibitem{Wilks} S.S. Wilks, {\em Mathematical Statistics} (John Wiley \& Sons, New York, 1962).

\bibitem{Fukunaga} K. Fukunaga, {\em Introduction to Statistical Pattern Recognition}
(Elsevier, New York, 1990).

\bibitem{Smith} For a nice pedagogical introduction to PCA, see
L.I. Smith, ``A tutorial on Principal Components Analysis" (2002).
%(csnet.otago.ac.nz/cosc453/student$_$tutorials/principal$_$components.pdf).

\bibitem{Wishart} J. Wishart, Biometrica {\bf 20}, 32 (1928).

\bibitem{Johnstone} I.M. Johnstone, Ann. Statist. {\bf 29}, 295 (2001).

\bibitem{Preisendorfer} R.W. Preisendorfer, {\em Principal Component Analysis in Meteorology
and Oceanography} (Elsevier, New York, 1988).

\bibitem{BP} J.-P. Bouchaud and M. Potters, {\em Theory of Financial Risks} (Cambridge University
Press, Cambridge, 2001).

\bibitem{Burda} Z. Burda and J. Jurkiewicz, Physica {\bf A 344}, 67 (2004);
Z. Burda, J. Jurkiewicz and B. Waclaw, Acta Physica Polonica {\bf
B 36}, 2641 (2005) and references therein.

%\bibitem{SP} A.M. Sengupta and P.P. Mitra, Phys. Rev. E {\bf 60}, 3389 (1999).

\bibitem{SP} E. Telatar, European Transactions on Telecommunications {\bf 10}(6), 585 (1999).

\bibitem{Fyo1} Y.V. Fyodorov and H.-J. Sommers, J. Math. Phys. {\bf 38}, 1918 (1997); Y.V. Fyodorov
and B.A. Khoruzhenko, Phys. Rev. Lett. {\bf 83}, 66 (1999).

\bibitem{QCD} E.V. Shuryak and J.J.M. Verbaarschot, Nucl. Phys. {\bf A}560, 306 (1993); J.J.M. Verbaarschot, Phys. Rev. Lett. {\bf 72}, 2531 (1994).

\bibitem{Johansson} K. Johansson, Comm. Math. Phys. {\bf 209}, 437 (2000).

\bibitem{MZ1} S. Maslov and Y.C. Zhang, Phys. Rev. Lett. {\bf 87}, 248701 (2001).

\bibitem{Z2} Y.K. Yu and Y.C. Zhang, Physica {\bf A312}, 1 (2002).


\bibitem{JN} R.A. Janik and M. A. Nowak, J. Phys. A: Math. Gen. {\bf 36}, 3629 (2003).

\bibitem{Dys:new} F.J. Dyson, J. Math. Phys. {\bf 3} 140 (1962).

\bibitem{MP} V.A. Marcenko and L.A. Pastur, Math. USSR-Sb, {\bf 1}, 457 (1967).

\bibitem{Dyson} F.J. Dyson, Rev. Mex. Fis., {\bf 20}, 231
(1971).

\bibitem{Bohigas} O. Bohigas and J. Flores, Rev. Mex. Fis.,
{\bf 20}, 217 (1971).

\bibitem{Edelman} A. Edelman, SIAM J. Matrix Anal. Appl. {\bf 9}, 543 (1988).

\bibitem{Forrester} P.J. Forrester, Nucl. Phys. {\bf B 402}, 709 (1993).

\bibitem{DE} I. Dumitriu and A. Edelman, J. Math. Phys. {\bf 43}, 5830 (2002);

\bibitem{ES} A. Edelman and B. Sutton, arXiv:math-ph/0607038.


\bibitem{DM} D.S. Dean and S.N. Majumdar, Phys. Rev. Lett. {\bf 97}, 160201 (2006).


\bibitem{Hiai} F. Hiai and D. Petz, {\em The semicircle law, free random variables and entropy} (American Mathematical Society, Providence,
2000).

\bibitem{Deift} P. Deift, A. Its and I. Krasovsky, arXiv:math.FA/0609451


\bibitem{Mehta} M.L. Mehta, Random Matrices, 3rd Edition,
(Elsevier-Academic Press) (2004).

\bibitem{Chen} Y. Chen and S.M. Manning, J. Phys. A: Math. Gen.
{\bf 29} 7561 (1996); {\em ibid} {\bf 27} 3615 (1994).


\bibitem{James} A.T. James, Ann. Math. Statistics {\bf 35}, 475 (1964).

\bibitem{Wigner}{E.P. Wigner, Proc. Cambridge Philos. Soc. {\bf 47},
790 (1951).}

\bibitem{TW1} C. Tracy and H. Widom, Comm. Math. Phys. {\bf 159}, 151
(1994); {\em ibid} {\bf 177}, 727 (1996); For a review see {\em
Proceedings of the International Congress of Mathematicians},
Beijing 2002, Vol. I, ed. LI Tatsien, Higher Education Press,
Beijing 2002, pgs. 587-596.








\bibitem{CGG} A. Cavagna, J.P. Garrahan, and I. Giardina, Phys. Rev. B {\bf 61}, 3960 (2000).

\bibitem{Fyodorov} Y.V. Fyodorov, Phys. Rev. Lett. {\bf 92}, 240601 (2004) ;
Acta Physica Polonica B {\bf 36}, 2699 (2005).

\bibitem{Susskind} L. Susskind, arXiv:hep-th/0302219; M.R. Douglas,
B. Shiffman, and S. Zelditch, Comm. Math. Phys. {\bf 252}, 325
(2004).

\bibitem{MH} L. Mersini-Houghton, Class. Quant. Grav. {\bf 22}, 3481 (2005).

\bibitem{AE} A. Azami and R. Easther, J. Cosmol. Astropart. Phys.
JCAP03 013 (2006).

\bibitem{BrayDean} A.J. Bray and D.S. Dean, arXiv:cond-mat/0611023.

\bibitem{FSW} Y.V. Fyodorov, H-J. Sommers and I. Williams, arXiv:cond-mat/0611585.

\bibitem{Kanz} E. Kanzieper, Phys. Rev. Lett. {\bf 89} 250201 (2002)

\bibitem{Tricomi1} F.G. Tricomi, {\em Integral Equations}, (Pure Appl. Math V, Interscience,
London, 1957).


\bibitem{Abramo} M. Abramowitz and I.A. Stegun, {\em Handbook of Mathematical
Functions} (Dover Publications,Inc.,New York 1972).







\end{thebibliography}
\end{document}